%% file: ACII_2021.tex
\documentclass[conference]{IEEEtran}
\pdfoutput=1
\IEEEoverridecommandlockouts
\usepackage{cite}
\usepackage{amsmath,amssymb,amsfonts}
\usepackage{algorithmic}
\usepackage{graphicx}
\usepackage{textcomp}
\usepackage{xcolor}
\usepackage{booktabs, makecell}
\usepackage{multirow}

\definecolor{darkblue}{RGB}{0,0,127}

\usepackage{hyperref}
\def\BibTeX{{\rm B\kern-.05em{\sc i\kern-.025em b}\kern-.08em
    T\kern-.1667em\lower.7ex\hbox{E}\kern-.125emX}}
    
\usepackage{fancyhdr}
\begin{document}

\title{How diverse is the ACII community? \\ Analysing gender, geographical and business diversity of Affective Computing research}

\author{\IEEEauthorblockN{Isabelle Hupont}
\IEEEauthorblockA{\textit{Joint Research Centre, European Commission} \\
Seville, Spain \\
isabelle.HUPONT-TORRES@ec.europa.eu}
\and
\IEEEauthorblockN{Songül Tolan}
\IEEEauthorblockA{\textit{Joint Research Centre, European Commission} \\
Seville, Spain  \\
songul.TOLAN@ec.europa.eu}
\and
\IEEEauthorblockN{Ana Freire}
\IEEEauthorblockA{\textit{Universitat Pompeu Fabra} \\
Barcelona, Spain \\
ana.freire@upf.edu}
\and
\IEEEauthorblockN{Lorenzo Porcaro}
\IEEEauthorblockA{\textit{Universitat Pompeu Fabra} \\
Barcelona, Spain  \\
lorenzo.porcaro@upf.edu}
\and
\IEEEauthorblockN{Sara Estevez}
\IEEEauthorblockA{\textit{Universitat Pompeu Fabra} \\
Barcelona, Spain  \\
sara.estevez02@estudiant.upf.edu}
\and
\IEEEauthorblockN{Emilia Gómez}
\IEEEauthorblockA{\textit{Joint Research Centre, European Commission} \\
Seville, Spain  \\
emilia.GOMEZ-GUTIERREZ@ec.europa.eu}
}

\maketitle
\thispagestyle{fancy}

\begin{abstract}
ACII is the premier international forum for presenting the latest research on affective computing. In this work, we monitor, quantify and reflect on the diversity in ACII conference across time by computing a set of indexes. We measure diversity in terms of gender, geographic location and academia vs research centres vs industry, and consider three different actors: authors, keynote speakers and organizers. Results raise awareness on the limited diversity in the field, in all studied facets, and compared to other AI conferences. While gender diversity is relatively high, equality is far from being reached. The community is dominated by European, Asian and North American researchers, leading the rest of continents under-represented. There is also a strong absence of companies and research centres focusing on applied research and products. This study fosters discussion in the community on the need for diversity and related challenges in terms of minimizing potential biases of the developed systems to the represented groups. We intend our paper to contribute with a first analysis to consider as a monitoring tool when implementing diversity initiatives. The data collected for this study are publicly released through the European divinAI initiative.
\end{abstract}


\input{Input/1_intro.tex}
\input{Input/2_related_work}
\input{Input/3_methodology}
\input{Input/4_Results}
\input{Input/5_conclusions}

\section*{Acknowledgment}

This work is partially supported by the European Commission under the TROMPA project (H2020 770376) and the HUMAINT project of the Joint Research Centre. 

\bibliographystyle{IEEEtran}
\bibliography{IEEEabrv,literature.bib}

\end{document}

%% file: Input/1_intro.tex
\section{Introduction}

Rosalind Picard first coined the term Affective Computing and presented its aims and visions in the late 90s~\cite{picard2000affective}. Since then, much effort has been placed on finding special features with which machines, computers and digital devices are enabled to identify and respond to users' emotions by means of machine learning processes.

The International Conference on Affective Computing and Intelligent Interaction (ACII) is widely acknowledged as the premier international forum for presenting the latest research on affective and multimodal human-machine interaction and systems. It is the most relevant source of conference publications in the field, complementing journal papers published at IEEE Transactions on Affective Computing~\cite{guo2020bibliometric}, and the most prominent forum for networking and research community building. The conference is celebrating its 9th edition in 2021, being held every two years since 2005 in geographical turns between USA-Europe-Asia. ACII 2021 has as special theme \textit{Ethical Affective Computing}, reflecting the recent trends in several interdisciplinary research communities to study the social and ethical implications of their research~\cite{gomez2018assessing}. The current edition considers related aspects such as fairness, transparency, diversity and bias in emotional artificial intelligence (AI) as its core topics for the first time. In this context, this paper focuses on the analysis of diversity in the ACII field.

ACII promotes research at the cross-roads between engineering and human sciences. It includes topics as varied as the study of psychology and cognition of affect in designing computational systems, emotion detection from different modalities (face, audio, gestures, physiology), synthesis of affect in virtual agents or social robots, literary, art and cinema affective studies, emotional e-learning systems, advertisement and other real-time applications, to name a few. Affective Computing is indeed a discipline-diverse field by nature.

However, the concept of diversity has many other facets apart from the interdisciplinarity view~\cite{stirling2007general}. In the context of this work, diversity refers to the existence of variations of different characteristics in a group of people, more particularly in a research community. These characteristics could be everything that makes each person unique, such as cognitive skills and personality traits, along with the factors that shape identity (e.g. race, age, gender, religion, cultural background).

Evidence suggests that diverse teams outperform homogeneous groups on complex tasks, including improved problem solving, increased innovation and more accurate predictions, all of which lead to better performance and results~\cite{swartz2019science,freeman2014collaboration,alshebli2018preeminence}. Diverse and inclusive scientific communities can generate new research questions not yet being asked in their particular discipline or culture, develop inclusive methodologies to better understand broader populations, and offer novel approaches to problem solving from multiple and different perspectives. Diverse groups have been shown to publish more articles, and these receive more citations per article~\cite{adams2013fourth}. Diversity thus enhances excellence, inclusion, generality and innovation.
Affective Computing is in the core of AI human-centered applications, having a strong social and ethical impact, dealing with applications with a strong impact on humans such as emotion recognition and induction~\cite{humaint}. In order for these applications to incorporate different views and to avoid biases and potential discrimination (e.g. with respect to gender, race, cultural background or languages), there is a need to ensure that research is carried out with a varied perspective, which should be reflected by the research community~\cite{fan2021demographic,rukavina2016affective}.

Beyond ACII, it is well recognized that the Artificial Intelligence (AI) field is facing a diversity crisis, being male-dominated and with little representation of people from least developed countries~\cite{freire2021measuring}. As a field related to AI~\cite{corea2019ai}, Affective Computing is likely to follow the same trend. However, to the best of our knowledge, diversity in the Affective Computing research community has not been fully quantified and studied to date. 

In this paper, we monitor, quantify and reflect on the diversity in ACII conference across time by computing a set of indexes. We measure diversity in terms of gender, geographical location and academia vs industry, and consider three different actors: authors, keynote speakers and organizers. Results are publicly released through the European divinAI initiative.

%% file: Input/2_related_work.tex
\section{Related work}

\subsection{Bibliometric studies on Affective Computing}

Some recent bibliometric studies have analyzed the past 20-25 years of publications in the field of Affective Computing, from 1995 to 2020~\cite{guo2020bibliometric,ho2021affective,pestana2018global}. They all follow a similar methodology, searching for publications using the keyword ``Affective Computing'' in Web-of-Science (WoS), an online subscription-based scientific citation indexing service. Then, they mostly identify major publications, leading journals, key research topics, and most productive authors, institutions and countries in terms of number of published papers. 

The growth rate of scientific production in the field differs slightly from one study to another, but is generally very high. According to Ho et al.~\cite{ho2021affective}, the annual growth rate of published papers is 12.5\%. Pestana et al.~\cite{pestana2018global} found that publications double in 4.02 years time. Thus, Affective Computing is attracting more and more researchers every year.

Although none of these studies performs an explicit analysis of scientific production in academia vs research centers vs industry, results show that top contributing authors and institutions in the field come exclusively from academia. They also agree on that most productive countries are USA, China, England and Germany. Ho et al.~\cite{ho2021affective} also studied collaboration networks by country. Interestingly, they found two major collaborative networks: the ``Asia Pacific cluster'' with USA, China, Singapore and Japan, and the ``European cluster'' with Germany, UK and Netherlands. While the cross-cultural diversity of the ``Asia Pacific cluster'' might bring a new wave of studies on the cross-cultural differences in emotionality and their implication for Affective Computing techniques, the ``European cluster'' risks falling into endogamy.

These bibliometric studies shine a light on the current status of Affective Computing research. However, they do not fully explore diversity in the field. For instance, they do not address gender (i.e. presence of female authors) and academia vs industry balance dimensions. Geographic diversity has been more studied, but in any case not sufficiently quantified with appropriate indexes or indicators.

\subsection{Biases in AI and Affective Computing}

Measuring, analyzing causes and effects, and mitigating biases in AI algorithms is a hot research topic nowadays. Buolamwini’s pioneer study on the ``gender shades'' of automated facial analysis~\cite{buolamwini2018gender} started a major movement on a large scale, raising alarms about the detrimental impact demographic (gender, age, ethnicity) AI systems' biases have in society. Since then, the AI community has devoted a lot of effort to reduce such biases, e.g. by creating more diverse datasets~\cite{karkkainen2019fairface,hupont2019demogpairs}, and by proposing new demographic-aware training methodologies~\cite{wang2020towards} and model architectures~\cite{ryu2017inclusivefacenet}. Even tech giants have already got on the demographic diversity train. For example, IBM has released ``Diversity in Faces''~\cite{merler2019diversity}, the largest labeled dataset containing about 1 Million images equally distributed across skin tones, genders and ages. 

The field of Affective Computing is not yet as active as other AI communities in the topic, but some prominent studies and tools have also been presented. For instance, Rukavina et al. studied how gender and age impact Affective Computing algorithms \cite{rukavina2016affective} and Fan et al. analysed demographic effects on facial emotion expression~\cite{fan2021demographic}. Another example is Emotion Miner~\cite{EmoMiner}, a data corpus by the company Neurodata Lab, containing affective data extracted from public videos across the globe, which can be adapted for any specific request, any cultural background or the needs of local communities.

Scientists are now concerned about demographic biases and the lack of diversity in AI systems, including Affective Computing systems, which are being well studied from the algorithmic perspective. However, there is a need for analysing diversity from another important tough widely understudied point of view: the scientific community itself, i.e., how diverse are researchers working in the field.

\subsection{Diversity in AI scientific communities}

Different reports, such as the Ethical guidelines for Trustworthy AI by the High-level Expert Group of the European Commission~\cite{EC_ethic_guidelines} and the last AI Now Institute report~\cite{west2019discriminating}, emphasize the urgency of fighting for diversity and reconsidering diversity in a broader sense, including gender, culture, origin and other attributes such as discipline or domain that can contribute to a more diverse research and development of AI systems.

However, little attention had been paid to the monitoring of diversity in AI to date. Only very recently, the DivinAI\footnote{DivinAI web platform: \url{https://divinai.org}} (Diversity in Artificial Intelligence) platform has been launched. It is an open and collaborative initiative promoted by the European Commission to research and develop a set of diversity indicators related to AI developments, with special focus on gender balance, geographical representation, and presence of academia vs companies~\cite{freire2021measuring}. These indicators are informative about the visibility of minorities in AI big events, minorities authorship in AI research contributions (conference proceedings), and minorities presence in AI organisation committees. The previous indicators are combined in order to assign one general diversity indicator to each big AI event. This can be useful to compare how different AI conferences care about minorities, to monitor the impact of diversity and inclusion policies (e.g. mentoring programs, travel grants) and to raise awareness on the need for more diverse research communities. 

The goal of DivinAI is addressed through a collaborative website in which anyone can contribute by adding data related to most relevant AI conferences worldwide, namely, details about keynotes speakers, members of organisation committee and authors. While the platform contains data about general AI conferences 
(e.g. ICML\footnote{Diversity dashboard for the International Conference on Machine Learning (ICML) in DivinAI: \url{https://divinai.org/conf/76/icml}}, IJCAI\footnote{Diversity dashboard for the International Joint Conference on AI (IJCAI) in DivinAI: \url{https://divinai.org/conf/77/ijcai}}, NeurIPS\footnote{Diversity dashboard for the Conference on Neural Information Processing Systems (NeurIPS) in DivinAI: \url{https://divinai.org/conf/78/neurips}}), it has never been used to quantify diversity in Affective Computing forums.

%% file: Input/3_methodology.tex
\section{Methodology}

To monitor diversity in the Affective Computing research community, we collected data from past ACII conferences. We followed and complemented the methodology proposed by the divinAI initiative~\cite{freire2021measuring}, thus allowing to compare with other related research communities and make the data and results publicly available in the platform.

\subsection{Description of diversity indexes}
\label{sec:div_idx}

The methodology proposes four diversity indexes: Gender Diversity Index (GDI), Geographic Diversity Index (GeoDI), Business Diversity Index (BDI) and overall Conference Diversity Index (CDI). They are grounded on Shannon~\cite{shannon1948mathematical} and Pielou~\cite{pielou1966measurement} indexes, which are widely used in the field of plant and animal ecology to measure biodiversity~\cite{mouillot1999comparison}. 

Shannon index $H'$ is defined as:
\begin{equation} 
    H'=-{\sum\limits_{i=1}^{S}}p_i\ln{p_i}
\end{equation}
\noindent where $p_i=n_i/N$ is the proportion of individuals from a given species $i$, i.e., the number of individuals from this species $n_i$ divided by the total number of individuals $N$; and $S$ is the number of different species. The Shannon index takes values between 1.5 and 3.5 in most biodiversity studies, and is rarely greater than 4. Note that it increases as both the richness and the evenness of the community increase. Richness describes the number of different species present in an area (the more species, the greater richness), while evenness refers to the relative abundance of the different species in an area (a similar abundance means more evenness).

The Pielou index $J'$ allows to compute Shannon evenness, discarding the richness factor:
\begin{equation} 
    J'={\frac{H'}{H'_{max}}}
\end{equation}
$H'_{max}$ is the maximum possible value of the Shannon index, which is given when all species are equally likely:
\begin{equation} 
    H'_{max}=-{\sum\limits_{i=1}^{S}}\frac{1}{S}\ln{\frac{1}{S}}=\ln{S}
\end{equation}
$J'$ values are in range $[0;1]$, 1 meaning the highest evenness.

Based on Shannon and Pielou indexes, the description of each diversity index as defined and computed in this work follows. Diversity indexes are computed separately for keynote speakers, conference organizers and paper authors, in order to study diversity connected to the different roles that community members have within a conference. They are later aggregated into a single indicator by average.\\

\noindent \textbf{Gender Diversity Index (GDI).}
We consider two different categories ($S=2$) in the gender dimension: ``male'' and ``female''. We compute Shannon evenness by means of the Pielou diversity index, which is calculated separately for the three different actors in the community:  keynote  speakers, authors and conference organisers. The final GDI is the average of the three Pielou indexes.  

\noindent \textbf{Business Diversity Index (BDI).}
We consider three different categories ($S=3$) in the business dimension: ``academia'', ``industry'' and ``research centre''. As for GDI, the final BDI is the average of the three actors' Pielou indexes. 

\noindent \textbf{Geographic Diversity Index (GeoDI).}
In this case we have multiple categories, being $S$ the number of countries represented by community actors. As we want to measure the richness together with the evenness, we compute the average of the three actors' Shannon indexes. Note that GDI and BDI are based on Pielou's index, and normalized in [0;1] by definition. As GeoDI is based on Shannon's index and thus not normalised, it is divided by 3.5, which is commonly approximated in the literature as an upper value for the Shannon formula~\cite{ibanez1995pedodiversity}. 
 
\noindent \textbf{Conference Diversity Index (CDI).}
Finally, the overall conference diversity index CDI is computed by averaging gender, geographic and business indexes: 
\begin{equation} 
    CDI=\frac{GDI+GeoDI+BDI}{3}
\end{equation}
\noindent \textbf{Interpretation of diversity indexes values.}
Our diversity indexes are all normalised in the range [0;1], higher values implying greater diversity. The main benefit of using these indexes (versus, for instance, percentage analyses per year, actor, gender, country or type of institution) is that they summarize in a single value the gender, geographical, business and overall diversity of a conference.

\subsection{Data collection and processing}

The data used to compute our diversity indicators are exclusively based on public domain information available at ACII conference proceedings. For each paper in a proceeding, we extracted its authors' names and individual affiliations at that time, and inferred gender, geographic and business information from them. We also collected the same data for keynote speakers and organizers\footnote{Organisers include: general chairs, program chairs,local organizing chairs, special session chairs, tutorial chairs, workshop and special track chairs, demo chairs, publication chairs, doctoral consortium chairs and industry chairs.}. Table~\ref{tab:dataset} shows the hosting country, number of papers, authors, keynote speakers and organisers for each of the 8 past ACII editions.

\begin{table}[hbt!]
    \centering
    \scalebox{0.9}{
    \begin{tabular}{cccccc}
        \toprule
        ACII ed. & Host & \#papers & \#authors & \#keynotes & \#organisers\\
        \midrule
             2005 & China & 126 & 304 & 4 & 11\\
             2007 & Portugal & 94 & 281 & 4 & 19\\
             2009 & Netherlands & 146 & 510 & 3 & 16\\
             2011 & USA & 141 & 431 & 3 & 20\\
             2013 & Switzerland & 160 & 556 & 3 & 15\\
             2015 & China & 155 & 620 & 3 & 20\\
             2017 & USA & 96 & 347 & 3 & 22\\
             2019 & UK & 108 & 472 & 3 & 36\\
        \bottomrule
        \\
    \end{tabular}
    }
    \caption{Number of papers, authors, keynote speakers and organisers from past ACII conferences analyzed in this study.}
    \label{tab:dataset}
\end{table}

Depending on the edition, ACII proceedings are issued by two different publishers: IEEEXplore (ACII 2009, 2013, 2015, 2017 and 2019) and Springer Lecture Notes in Computer Science (ACII 2005, 2007 and 2011). The first step was to collect the name and affiliations of all conference actors. For keynote speakers and organisers, data were manually extracted from proceedings' foreword pages. The extraction of authors' names and affiliations was performed in an automatic or semi-automatic manner, depending on the publisher. In the case of IEEEXplore, the process could be fully automatised thanks to the search and CSV export tools available in its web platform. 
For Springer proceedings, we used DBLP search and its XML export tools 
to help automatise the process, but unfortunately affiliations were missing for about 80\% of the authors. We then had to complete manually remaining affiliations through manual web search at SpringerLink. 
In all cases, duplicated authors (i.e. authors that contributed to several papers in the same proceedings) were considered only once.

The second step was to infer gender from first names and surnames, as proceedings do not include gender information. The NamSor\footnote{NamSor web page: \url{https://www.namsor.com/}} gender classifier library was used for that purpose. Nationality is also not included in conference proceedings, the only information available being the country of affiliation. Thus, we considered the affiliation country of each individual at the conference time for computing GeoDI.

Finally, affiliation was also used to determine whether conference participants belonged to industry, academia or research centres. The process was semi-automated by searching keywords such as ``university'', ``college'', ``Inc.'', ``Corp'', ``GmbH'' or ``centre'', and then refining the results manually (in some specific cases, web search was needed). All the data collected in this study are publicly available, so that it can be reproduced and extended (c.f. Section~\ref{sec:divinAI}).

%% file: Input/4_results.tex
\section{Results and comparison with other conferences}

\begin{figure}[thb!]
    \centering
    \includegraphics[width=0.9\linewidth]{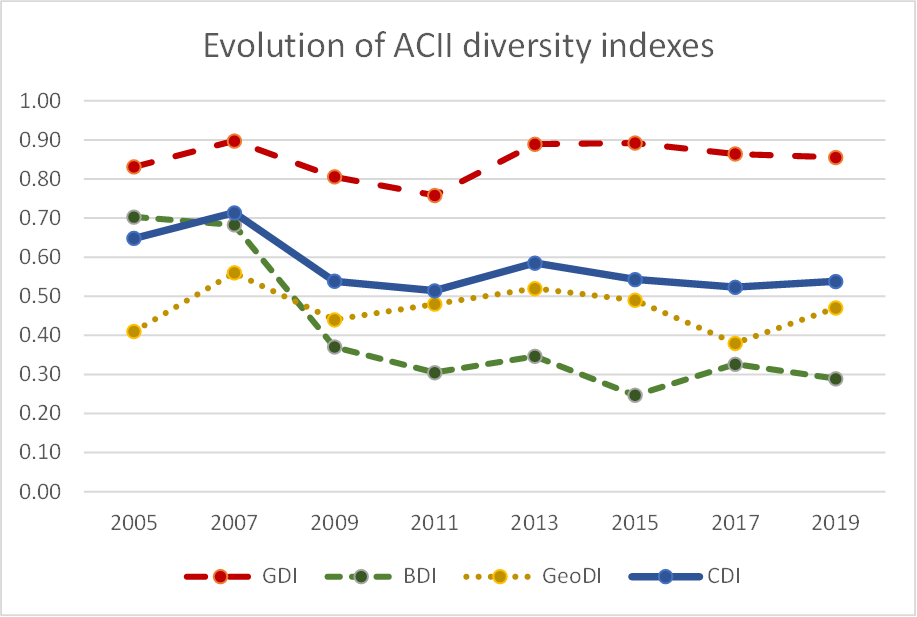}
    \caption{Evolution of ACII diversity indexes from 2005 to 2019 editions. Best viewed in color.}
    \label{fig:evol_acii_idx}
\end{figure}

\begin{figure}[htb!]
    \centering
    \includegraphics[width=0.87\linewidth]{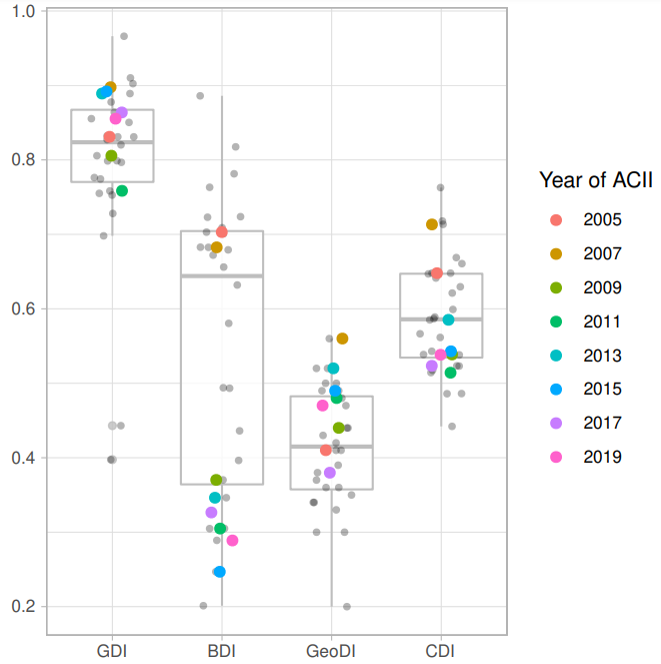}
    \caption{Boxplot distribution of diversity indexes' values across AI conferences. ACII data are represented with colored circles. Best viewed in color.}
    \label{fig:boxplots}
\end{figure}

Collected gender, geographic and business information has been used to compute the diversity indexes presented in Section~\ref{sec:div_idx} for the 8 past ACII editions. Figure~\ref{fig:evol_acii_idx}, Table~\ref{tab:overall_results} and Figure~\ref{fig:boxplots} show obtained results, which are discussed and compared with other relevant AI conferences in next sections.

\begin{table}[tbh!]
    \centering
    \scalebox{0.9}{
    \begin{tabular}{lcccccccc}
        \toprule
        Conference & GDI & BDI & GeoDI & CDI\\
        \midrule
             ACII 2005 & 0.83 & 0.70 & 0.41 & 0.65 \\
             ACII 2007 & 0.90 & 0.68 & \textbf{0.56} & 0.71 \\
             ACII 2009 & 0.81 & 0.37 & 0.44 & 0.54 \\
             ACII 2011 & 0.76 & 0.30 & 0.48 & 0.51 \\
             ACII 2013 & 0.89 & 0.35 & \underline{0.52} & 0.59 \\
             ACII 2015 & 0.89 & 0.25 & 0.49 & 0.54 \\
             ACII 2017 & 0.86 & 0.33 & 0.38 & 0.52 \\
             ACII 2019 & 0.86 & 0.29 & 0.47 & 0.54 \\
        \midrule
         NeurIPS 2016 & 0.75 & 0.63 & 0.30 & 0.56 \\
         NeurIPS 2017 & 0.73 & 0.67 & 0.36 & 0.59 \\
         NeurIPS 2018 & 0.70 & 0.71 & 0.36 & 0.59 \\
         NeurIPS 2019 & 0.88 & 0.72 & 0.34 & 0.65 \\
         NeurIPS 2020 & 0.90 & \textbf{0.89} & 0.50 & \textbf{0.76} \\
        \midrule
           IJCAI 2017 & 0.77 & 0.30 & 0.49 & 0.52 \\
           IJCAI 2018 & 0.80 & 0.40 & 0.44 & 0.55\\
           IJCAI 2019 & 0.83 & 0.20 & 0.43 & 0.49 \\
        \midrule
            ICML 2017 & 0.76 & 0.72 & 0.41 & 0.63 \\
            ICML 2018 & 0.78 & 0.44 & 0.34 & 0.52 \\           
            ICML 2019 & 0.80 & 0.70 & 0.30 & 0.60 \\   
            ICML 2020 & 0.83 & 0.78 & 0.33 & 0.65 \\            
        \midrule
            ECAI 2020 & \underline{0.91} & 0.49 & \underline{0.52} & 0.64 \\
        \midrule
       ACM FAccT 2020 & \textbf{0.97} & \underline{0.82} & 0.37 & \underline{0.72} \\ 
        \bottomrule
        \\
    \end{tabular}
    }
    \caption{Top rows: ACII diversity indexes. Bottom rows: Indexes of other relevant AI conferences, that had been introduced in divinAI prior to this work. Best diversity values are marked in bold, second best underlined.}
    \label{tab:overall_results}
\end{table}

\subsection{Gender diversity}

In general, values obtained for GDI are high, ranging from 0.76 (2011) to 0.90 (2007). In the recent years (2017-2020), they are in line with and even surpass some general AI conferences such as NeurIPS, IJCAI and ICML. They are however far from the value of 0.97 obtained by ACM FAccT (ACM Conference on Fairness, Accountability and Transparency), a conference paying particular attention to gender diversity.

\begin{figure}[htb!]
    \centering
    \includegraphics[width=0.9\linewidth]{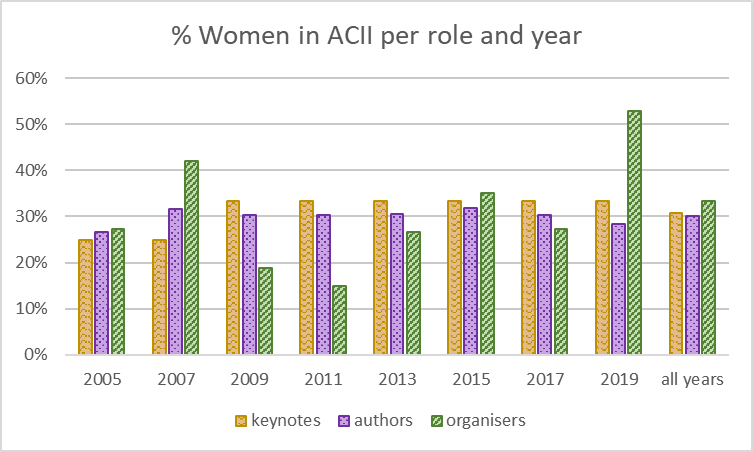}
    \caption{Percentage of women in ACII conferences, per role (keynote, author, organiser) and edition. Best viewed in color.}
    \label{fig:perc_women_kao}
\end{figure}

Figure~\ref{fig:perc_women_kao} shows the percentage of women, depending on their role in the conference (keynotes, authors, organisers) and per year. As it can be seen, the equality threshold of 50\% is never reached by any category (excepting for organisers in 2019), meaning that an effort is still to be done to promote gender balance in ACII conferences. Nevertheless, authors and keynotes maintain acceptable percentages across editions, with a women representation of about 30\%. Every past edition has had one female keynote speaker, the rest being males. In the last edition (2019), the representation of women in the organising committee has notably increased, which would be a very good practice to follow for future editions, as they were underrepresented in the past.

\begin{figure}[htb!]
    \centering
    \includegraphics[width=0.9\linewidth]{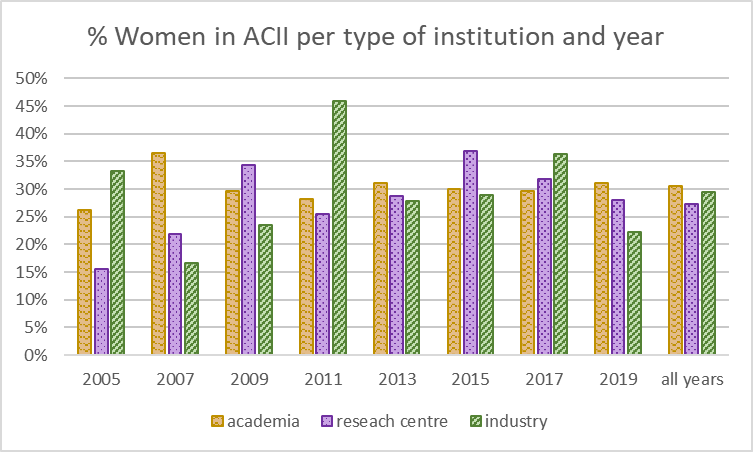}
    \caption{Percentage of women in ACII conferences, per type of institution (academia, research centre or industry) and edition. Best viewed in color.}
    \label{fig:perc_women_affi}
\end{figure}

The presence of ACII women in academia vs research centres vs industry is illustrated in Figure~\ref{fig:perc_women_affi}. Again, the percentage of women for each type of institution is far below the 50\% equality threshold. However, they are represented above 20\% in most cases, thus no strong under-representation is found, even for industry. This is partly due to the prominent contribution of ACII's top 5 female researchers and their teams, which are listed in Table~\ref{tab:female_top}.

\begin{table*}
\centering
\scalebox{0.85}{
\begin{tabular}{llcclc}
\toprule
Name                             & Institution               & Keynote               & Organiser                   & Author (\#papers)                                   & Country                  \\
\toprule
\multirow{2}{*}{Rosalind Picard} & Industry (Affectiva)      & \multirow{2}{*}{2011} & 2005, 2007                  & 2005(1), 2007(1), 2009(4), 2011(1),                 & \multirow{2}{*}{USA}     \\
                                 & and Academia (MIT)        &                       & 2019                        & 2013(3), 2015(3), 2017(4), 2019(3)                  &          \\
\midrule
\multirow{2}{*}{Maja Pantic}     & Academia                  & \multirow{2}{*}{2015} & \multirow{2}{*}{2009, 2013} & \multirow{2}{*}{2009(2), 2011(1), 2013(1), 2015(4)} & \multirow{2}{*}{UK}      \\
                                 & (Imperial College London) &                       &                             &                                                     &          \\
\midrule
Nadia                            & Academia                  & \multirow{2}{*}{-}    & 2009, 2013                  & 2005(1), 2007(2), 2009(2), 2011(2),                 & \multirow{2}{*}{UK}      \\
Bianchi-Berthouze                & (Univ. College London)    &                       & 2017, 2019                  & 2013(5), 2015(4), 2017(1), 2019(4)                  &                          \\
\midrule
Catherine                        & Research centre           & \multirow{2}{*}{-}    & 2009, 2013                  & 2005(1), 2007(3), 2009(2), 2011(2),                 & \multirow{2}{*}{France}  \\
Pelachaud                        & (CNRS)                    &                       & 2019                        & 2013(2), 2015(5), 2019(2)                           &                          \\
\midrule
Ginevra                          & Academia                  & \multirow{2}{*}{-}    & \multirow{2}{*}{2011, 2019} & \multirow{2}{*}{2007(2), 2009(1), 2011(1), 2013(3)} & \multirow{2}{*}{Sweden}  \\
Castellano                       & (Uppsala Univ.)           &                       &                             &                                                     &       \\
\bottomrule
\\
\end{tabular}
}
\caption{ACII's top 5 female contributors. The years in which they participated as keynote speakers, conference organisers and authors are shown, together with the number of published papers per year (in brackets), their type of institution and working country.}
\label{tab:female_top}
\end{table*}

\subsection{Geographic diversity}

ACII is leading the board of AI conferences in terms of GDI index, with values ranging form 0.38 (2017) to 0.56 (2007). Throughout all editions, 50 different countries have participated in the conference. The GDI is quite constant across years (c.f. Figure~\ref{fig:evol_acii_idx}), thus there seems to be no significant effect of the location of the conference into diversity. 

\begin{figure}[htb!]
    \centering
    \includegraphics[width=0.9\linewidth]{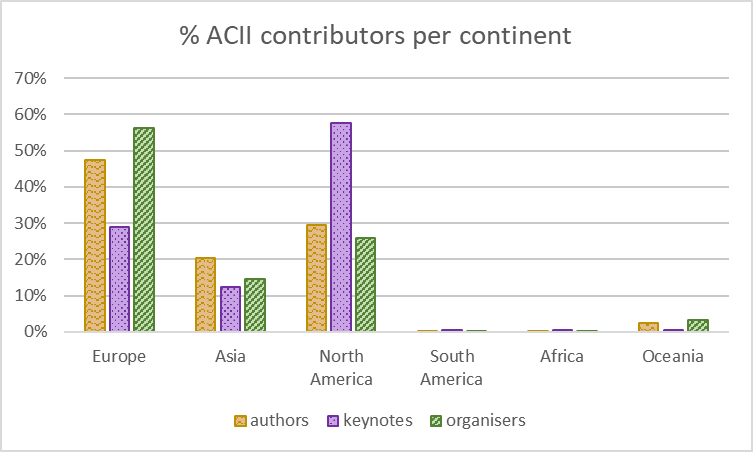}
    \caption{Percentage of authors, keynote speakers and organisers per continent (all years). Best viewed in color.}
    \label{fig:perc_cont}
\end{figure}

However, by looking at contributions per continent (Figure~\ref{fig:perc_cont}) it can be seen that Europe, Asia and North America are at the forefront; Oceania, South America and Africa being strongly under-represented. Indeed, the three leading continents are the ones hosting the conference. It is the right moment to foster the participation of other continents, and especially of least developed countries, as online post-pandemic events do not imply anymore expensive trips.

It is also interesting to highlight that most keynote speakers come from North America (mostly USA - 58\%), while Europeans are more wiling to organise the conference (56\%). Regarding authorship, the highest percentage of authors comes from Europe (47\%).

\subsection{Business diversity}

\begin{figure}[htb!]
    \centering
    \includegraphics[width=0.9\linewidth]{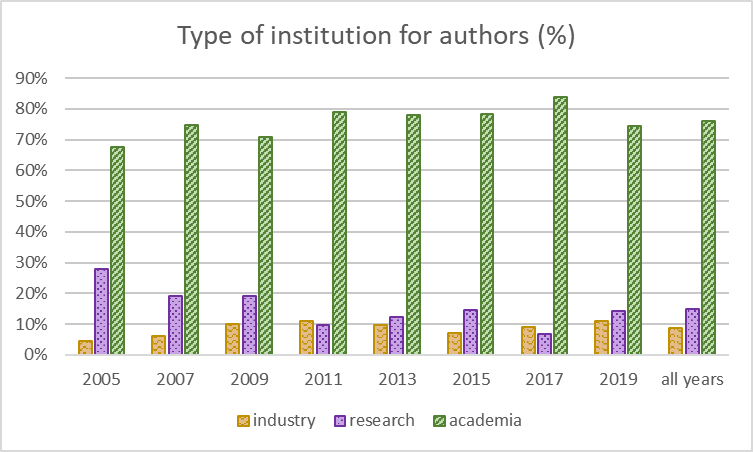}\\
     \includegraphics[width=0.9\linewidth]{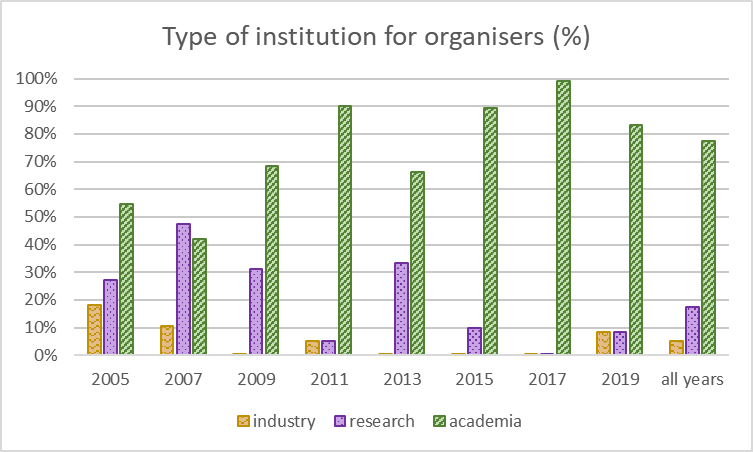}
    \caption{Distribution in percentage of the type of institution for authors and organisers, respectively, per year. Keynotes graph has not been included, due to their low representation in terms of numbers. Best viewed in color.}
    \label{fig:perc_business}
\end{figure}

ACII's BDI indexes are very low, especially in the six past editions where it ranges from 0.25 (2015) to 0.37 (2009). When compared to general AI conferences, it is far below NeurIPS, ICML and ACM FAccT. If we go deeper into the percentages shown in Figure~\ref{fig:perc_business}, we can see that the main reason is the high imbalance with a strong presence of academia vs research centres and industry. The imbalance ratios for organisers are around 1:15 and 1:4 for industry and research centres, respectively; 1:8 and 1:5 for authors; and 1:6 for keynotes, both for industry and research centres. 

A possible explanation for the low presence of industry and research centres might be that these kind of institutions focus more on applied research and final products. However, some Affective Computing technologies and applications have not yet reached enough technological maturity to go to the market~\cite{ho2021affective}. Ethical concerns are particularly slowing down the market penetration of Affective Computing products.

There are however prominent companies in the ACII community. Top 5 are listed in Table~\ref{tab:industry_top}. As it can be seen, 3 of them are very large companies: Philips, NTT and Microsoft. Remaining two, Affectiva and AudEERING, are startups (now SMEs) founded by two top ACII former academia authors: Rosalind Picard (Massachusetts Institute of Technology - MIT) and Bj\"orn Schuller (Technical University of Munich - TUM).

An effort has to be made to involve more company researchers in the conference organisation and to attract more company authors, perhaps through collaborations with universities and, especially, research centres. Also it would be necessary to foster more discussion about real-world applications, social impact and ethical concerns incorporating the industrial view. This could help to the creation of more startups and knowledge-transfer initiatives in the community.

\begin{table}
\centering
\scalebox{0.85}{
\setlength{\tabcolsep}{0.1\tabcolsep}
\begin{tabular}{lcccccc}
\toprule
\multirow{2}{*}{Name}      & Type of                & \multirow{2}{*}{Country} & Author                            & \multirow{2}{*}{Keynote} & \multirow{2}{*}{Organiser} & \multirow{2}{*}{Sponsor}  \\
                           & company                &                          & (\#papers)                        &                          &                            &                           \\
\toprule
\multirow{2}{*}{Philips}   & \multirow{2}{*}{Large} & \multirow{2}{*}{Netherlands}  & 2009(7)              & \multirow{2}{*}{-}       & \multirow{2}{*}{-}                            & \multirow{2}{*}{2009}               \\
                           &                        &                               & 2011(2)              &    &  &  \\
\midrule
\multirow{2}{*}{NTT}       & \multirow{2}{*}{Large} & \multirow{2}{*}{Japan}   & 2013(1), 2017(2)                 & \multirow{2}{*}{-}       & \multirow{2}{*}{2019}      & \multirow{2}{*}{-} \\
                           &                        &                          & 2019(3)                           &                          &                            &                           \\
\midrule
\multirow{2}{*}{Microsoft} & \multirow{2}{*}{Large} & \multirow{2}{*}{USA}     & 2013(2), 2017(1)                 & -                        & \multirow{2}{*}{2019}      & 2007, 2017         \\
                           &                        &                          & 2013(2)                           &                          &                            & 2019                      \\
\midrule
\multirow{2}{*}{Affectiva} & SME                    & \multirow{2}{*}{USA}     & 2013(1), 2015(1)                 & \multirow{2}{*}{2011}    & \multirow{2}{*}{2019}      & \multirow{2}{*}{-}        \\
                           & (MIT startup)          &                          & 2019(1)                           &                          &                            &                           \\
\midrule
\multirow{2}{*}{audEERING} & SME                    & \multirow{2}{*}{Germany} & 2015(8)  & \multirow{2}{*}{-}       & 2013, 2015                & \multirow{2}{*}{2017}     \\
                           & (TUM startup)          &                          &           2017(1)                        &                          & 2017, 2019                 &       \\                   
\bottomrule
\\
\end{tabular}
}
\caption{ACII's top 5 industry contributors. Years in which each company contributed as author, keynote, organiser and sponsor are provided, as well as the number of papers (in brackets), company size and headquarters country. }
\label{tab:industry_top}
\end{table}

\subsection{Overall conference diversity}

CDI compiles, in one single metric, all the diversity information discussed in the previous sections at a glance. Overall, ACII CDI indexes are close, and even surpass for some years, the ones of general AI conferences such as IJCAI, ICML, ECAI and NeurIPS. Again, it is far below ACM FAccT, as a reference conference in terms of diversity and inclusion. 

Perhaps most alarming is the lack of evolution of the CDI index, and of all other indexes, across ACII editions (Figure~\ref{fig:evol_acii_idx}). This stagnation might indicate the lack of specific policies taken to promote diversity, such as mentoring programs, visibility efforts, travel grants, committee diversity chairs and special workshops. 

\begin{figure}[tbh!]
    \centering
    \includegraphics[width=0.9\linewidth]{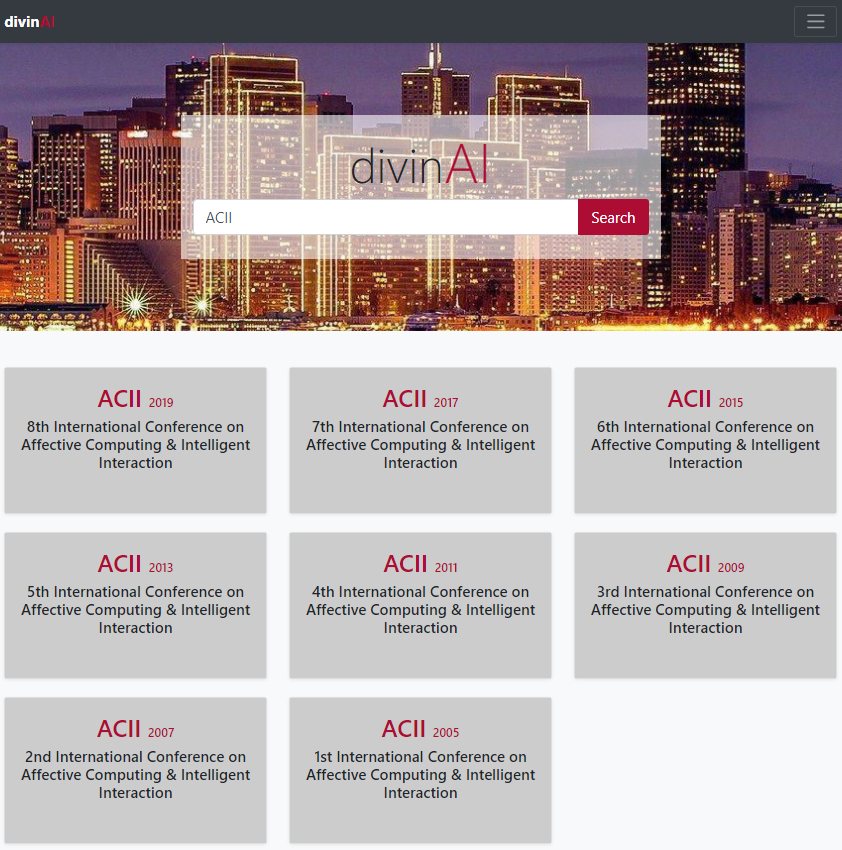}
    \caption{DivinAI's main menu for ACII conferences.} 
    \label{fig:divinAI_web}
\end{figure}

\subsection{Contribution to the open divinAI initiative}
\label{sec:divinAI}

Data collected in this work are publicly available at the divinAI web platform, being the only requirement for accessing them to freely register to the platform. This study can be reproduced and extended with data from future ACII editions, to monitor the impact of future initiatives at the community level. However, given the fact that we are dealing with sensitive personal data, we do not openly publish associations between name and gender information but we anonymised the data to be used for statistical purposes. For the sake of reproducibility, we keep all the data in our personal servers\footnote{The data protection procedure was reviewed by the UPF ethical committee in the context of the TROMPA project: \url{https://trompamusic.eu/}.}.

Figure~\ref{fig:divinAI_web} depicts divinAI's web interface. By searching ``ACII'', a main menu appears with links to the 8 past ACII editions. Each link points to the diversity dashboard of its corresponding edition. The main dashboard shows the values of the four diversity indexes. Then, other visualisations and statistics are also available by scrolling down, namely: histograms with gender and business percentages; interactive world maps with the geographic distribution of authors, keynote speakers and organisers; a timeline showing the evolution of ACII's CDI index across years; and a boxplot positioning the chosen ACII edition with regards to other AI conferences registered in the system. There is also an ``edit'' icon to modify in real-time the raw data used to compute indexes and statistics.



%% file: Input/5_conclusions.tex
\section{Conclusions}

This work has studied gender, geographical and business diversity of Affective Computing research, with the aim of raising awareness and discussion in the community on the need for diversity, the current limited diversity of ACII conferences and the challenges it brings in terms of minimizing potential biases of the developed systems and methodologies to the represented groups. We also intend our paper to contribute with a first analysis to consider as a monitoring tool when implementing diversity initiatives.

Our study has however some limitations. First, there are other relevant diversity dimensions. In the context of anti-discrimination and for the purpose of inclusive research and innovation, dimensions such as race, sexual orientation, religion or disability are at least equally important. However, there is no ethical basis for the inference of these dimensions based on name and affiliations. We note that the gender and nationality inference, may not be fully adequate either although we want to highlight the importance of monitoring the diversity of research communities in general. Moreover, gender had to be limited to a binary categorisation and may be misaligned with individual identification. Some of these issues could be solved if conference organizers collected more data at registration time, always ensuring strict personal data protection and governance rules.

As a future work, we plan to extend our diversity study by including data from the IEEE Transactions on Affective Computing journal, the first source of publications in Affective Computing. We also aim to explore more facets of diversity: authors' main discipline (e.g. Psychology, Neuroscience, Computer Science) or research topics (e.g. emotion synthesis, recognition, applications).
Finally, as part of the Affective Computing community, we would like to foster different initiatives for increasing diversity. This kind of actions have successfully started to be undertaken in other conferences\footnote{See, for instance, the activities launched by the Women in Machine Learning  \url{https://wimlworkshop.org} and Women in Music Information Retrieval Communities \url{https://wimir.wordpress.com/}}.